\documentclass[twocolumn,amsmath,amssymb,prl,groupedaddress,nobibnotes,showpacs]{revtex4}

\usepackage{graphicx}

\begin{document}

\title{Nonlocal Dispersion Cancellation using Entangled Photons}

\author{So-Young Baek}\email{simply@postech.ac.kr}\affiliation{Department of Physics, Pohang University of Science and
Technology (POSTECH), Pohang, 790-784, Korea}

\author{Young-Wook Cho}
\affiliation{Department of Physics, Pohang University of Science and
Technology (POSTECH), Pohang, 790-784, Korea}

\author{Yoon-Ho Kim} \email{yoonho@postech.ac.kr} \affiliation{Department of Physics,
Pohang University of Science and Technology (POSTECH), Pohang,
790-784, Korea}

\date{\today}

\begin{abstract}
A pair of optical pulses traveling through two dispersive media will become broadened and, as a result, the degree of coincidence between the optical pulses will be reduced. For a pair of entangled photons, however, nonlocal dispersion cancellation in which the dispersion experienced by one photon cancels the dispersion experienced by the other photon is possible. In this paper, we report an experimental demonstration of nonlocal dispersion cancellation using entangled photons. The degree of two-photon coincidence is shown to increase beyond the limit attainable without entanglement. Our results have important applications in fiber-based quantum communication and quantum metrology.
\end{abstract}


\maketitle




\section{Introduction}

Consider an ultrafast optical pulse propagating through a dispersive
media. Due to the group velocity dispersion, the wave packet of the
pulse will get broadened. When two such pulses, initially coincident
in time, travel through two different dispersive media, each pulse
will experience dispersion independently of the other. The
dispersive broadening of the two wave packets will then reduce
degree of temporal coincidence.

When a pair of entangled photons is considered instead of two
classical ultrafast pulses, a surprising result can occur: the
dispersion experienced by one photon can be canceled by the
dispersion experienced by the other and the dispersion cancellation
is independent of the distance between the two photons. The nonlocal
dispersion cancellation effect, originally proposed in
Ref.~\cite{franson}, is a further manifestation of the nonlocal
nature of quantum theory and is of importance in photonic quantum
information where entangled photons are distributed through
dispersive media. For instance, consider the Hong-Ou-Mandel-type
quantum interference effect resulting from the overlap between
nonclassical photonic wavepackets \cite{hom}. The temporal mode mismatch caused by the
group velocity dispersion effects of the media will severely reduce
the quality of quantum interference which in turn will have negative
impact on the efficiency of Bell-state measurement in quantum
teleportation \cite{tele}, the fidelity of the photonic quantum
gates \cite{rohde}, the quality of entanglement swapping
\cite{jennewein}, etc. The nonlocal dispersion cancellation effect
can be effectively utilized in these applications to remove unwanted
dispersive effects on nonclassical wave packets without the loss of
photons.

Although the nonlocal dispersion cancellation effect has been
theoretically shown to be important in many quantum metrology
applications \cite{giov01a,giov01b,fitch}, it has not been
conclusively demonstrated to date \cite{gis,gis2}. In this paper, we
report an explicit experimental demonstration of the nonlocal
dispersion cancellation effect using a pair of entangled photons. It
is important to note that the nonlocal dispersion cancellation of
Ref.~\cite{franson}, which we demonstrate in this paper, is
different from the dispersion cancellation effect of
Ref.~\cite{steinberg,nasr}, which is local
\cite{mention}.

\section{Theory}
\begin{figure}[t]
\centering
\includegraphics[width=3.0in]{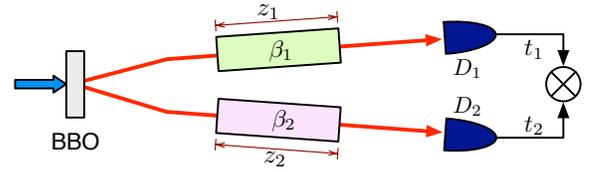}
\caption{Each photon of the entangled photon pair is subject to
different dispersion $\beta_1$ and $\beta_2$. The coincidence
circuit measures $G^{(2)}(t_1-t_2)$.}\label{idea}
\end{figure}
Let us first briefly discuss the theory behind nonlocal dispersion
cancellation \cite{franson,fitch}. Consider a pair of spontaneous
parametric down-conversion (SPDC) photons generated at a BBO crystal
pumped by a monochromatic pump, see Fig.~\ref{idea}. The quantum
state of the photon pair,
\begin{equation}\label{eq1}
|\psi\rangle = \int d\omega_1 d\omega_2
\mathcal{S}(\omega_1,\omega_2) a^\dagger(\omega_1)
a^\dagger(\omega_2)|0\rangle,
\end{equation}
is an entangled state since the two-photon joint spectral amplitude
$\mathcal{S}$ is non-factorizable, i.e.,
$\mathcal{S}(\omega_1,\omega_2) \neq \mathcal{S}(\omega_1)
\mathcal{S}(\omega_2)$, due to the monochromatic nature of the pump
\cite{kim05,baek08a}. The joint detection probability of the two
detectors $D_1$ and $D_2$ is proportional to the Glauber
second-order correlation function
\begin{equation}\label{gl1}
G^{(2)}(t_1,t_2) = |\langle0|E_{2}^{(+)}(t_{2})E_{1}^{(+)}(t_{1})|\psi\rangle|^2,
\end{equation}
where $E_{1}^{(+)}(t_{1})= \int d \omega \,
a(\omega_{1})e^{i\{k_{1}z_{1}-\omega_1 t_{1}\}}$, the positive
frequency part of the electric field operator at detector $D_1$, and
$E_2^{(+)}$ is defined similarly. Since each photon has spectral
distribution centered at certain central frequency, $\Omega_1$ and
$\Omega_2$, and assuming that the photons propagate through
dispersive media of length $z_1$ and $z_2$ as shown in
Fig.~\ref{idea}, the wave number of the photon can then be expressed
as $k_{i}(\Omega_{i}  \pm \nu) = k_{i}(\Omega_{i}) \pm \alpha_{i}\nu
+\beta_{i}\nu^{2}$ ($i=1,2$). Here $\nu$ is the detuning frequency
from the central frequency, and $\alpha$ and $\beta$ are the
first-order and the second-order dispersion which are responsible
for the wave packet delay and the wave packet broadening,
respectively.

For a monochromatic pump, the quantum state in eq.~(\ref{eq1}) can be re-written as
\begin{equation}\label{eq33}
|\psi\rangle = \int d\nu \mathcal{S}(\nu ) a^\dagger(\Omega_1+\nu) a^\dagger(\Omega_2-\nu)|0\rangle,
\end{equation}
with $\mathcal{S}(\nu )=\textrm{sinc}(\nu D L/2)$ for nondegenerate type-I SPDC. Here, $L$ and $D \equiv 1/u_2 - 1/u_1$ are the BBO crystal thickness and the inverse group velocity difference between the photon pair in the BBO crystal, respectively. Equation (\ref{gl1}) can then be expressed as,
\begin{eqnarray}\label{eq5}
 \nonumber G^{(2)}(t_{1}-t_{2}) &=&   \left| \int_{-\infty}^{\infty}d \nu \:
\mathcal{S}(\nu)\: e^{i\nu(t_{1}-t_{2})}\right.
\left. \times
e^{i(\alpha_{1}z_{1}-\alpha_{2}z_{2})\nu}e^{i(\beta_{1}z_{1}+\beta_{2}z_{2})\nu^2}
\right|^2.
\end{eqnarray}
The above expression can be analytically evaluated by approximating the joint spectral amplitude as a Gaussian function $\mathcal{S}(\nu) \approx e^{-\gamma (\nu D L)^2}$. The value $\gamma=0.04822$ was chosen so that the approximated Gaussian function would have the same full width at half maximum as the original function.

We, therefore, arrive at
\begin{eqnarray}\label{eq6}
G^{(2)}(t_{1}-t_{2}) \approx C e^{-(t_{1}-t_{2}-\bar{\tau})^2/2\sigma^2},
\end{eqnarray}
where $\bar{\tau}=\alpha_{2}z_{2}-\alpha_{1}z_{1}$ and $\sigma^2=\gamma D^{2} L^{2}+(\beta_{1}z_{1}+\beta_{2}z_{2})^{2}/\gamma D^{2} L^{2}$ are the overall time delay between the signal and the idler photons and the width of $G^{(2)}$ function after propagating through the dispersion media, respectively. Note that $C=\pi/\sqrt{\gamma^{2}D^{4}L^{4}+(\beta_{1}z_{1}+\beta_{2}z_{2})^{2}}$ is an unimportant proportionality factor.

Finally, the full width at half maximum value of the temporal width of $G^{(2)}$ function is given as \cite{note0}
\begin{equation}\label{eq8}
\Delta t \approx2\sqrt{\frac{2\ln2}{\gamma D^{2}
L^{2}}}(\beta_{1}z_{1}+\beta_{2}z_{2}).
\end{equation}
The above equation shows precisely the effect we have been looking for: a positive dispersion $\beta_1$ experienced by photon 1 can be cancelled by a negative dispersion $\beta_2$ experienced by photon 2. Such nonlocal dispersion cancellation effect is shown to occur only if photon 1 and photon 2 are in a specific entangled state \cite{franson}.

\section{Experiment}

\begin{figure}[t]
\centering
\includegraphics[width=3.0in]{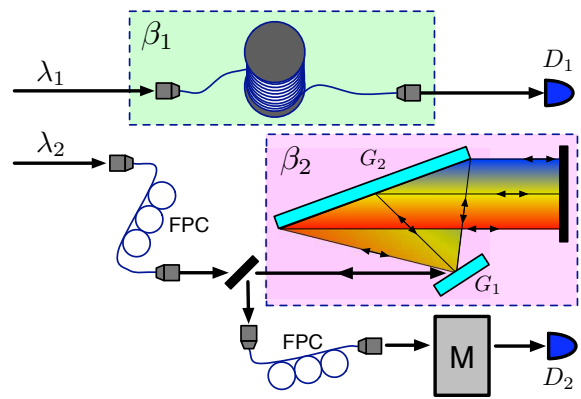}
\caption{Schematic of the experiment. Positive dispersion $\beta_1$ is introduced by a 1.6 km long single-mode optical fiber and negative dispersion $\beta_2$ is introduced by using a pair of gratings ($G_1$ and $G_2$) and a mirror. FPC and M represent a fiber polarization controller and a monochromator, respectively. }\label{setup}
\end{figure}

The experimental setup to implement the nonlocal dispersion cancellation effect discussed above is schematically shown in Fig.~\ref{setup}. We first describe the entangled photon generation part of the experimental setup which is not shown in the figure. A 3 mm thick type-I $\beta$-barium borate (BBO) crystal is pumped by a cw diode laser operating at 408.2 nm (40 mW), generating a pair of collinear frequency-nondegenerate entangled photons centered at 896 nm and 750 nm via the SPDC process. Type-I collinear phase matching ensures that both the signal and idler photons of SPDC have very broad spectra centered  at 896 nm  (over 28 nm at FWHM) and 750 nm  (over 20 nm at FWHM), respectively. The photon pair, however, is strongly \textit{quantum} correlated in wavelength: if the signal photon is found to have the wavelength $\lambda_1$, then its conjugate idler photon will be found to have the wavelength $\lambda_2=\lambda_p \lambda_1/(\lambda_1-\lambda_p)$, where $\lambda_p =408.2$ nm. The photon pair is in the two-photon quantum superposition  (i.e., entangled) state of eq.~(\ref{eq33}).

To efficiently couple these photons into single-mode optical fibers, the pump laser was focused with a $f=300$ mm lens and the SPDC photons were coupled into single-mode optical fibers using $\times 10$ objective lenses located at 600 mm from the crystal. The pump beam waist at the focus is roughly 80 $\mu$m. The co-propagating photons were then separated spatially by using a beam splitter and two interference filters with 100 nm FWHM bandwidth, each centered at 896 nm and 750 nm. The 896 nm centered signal photon ($\lambda_1$ in Fig.~\ref{setup}) is then coupled into a 1.6 km long single-mode optical fiber which
introduces positive dispersion $\beta_1$ to the photon. As we shall show later, the effect of a
positive dispersion material to the entangled photon is to broaden
the biphoton wave packet \cite{valencia,baek08b,baek08c}.

To demonstrate the nonlocal dispersion cancellation effect, it is
necessary to introduce negative dispersion to the idler photon
($\lambda_2$ in Fig.~\ref{setup}). Among many potential methods for
introducing negative  dispersion \cite{book}, methods based
on a grating pair or a prism pair are often used in ultrafast optics
\cite{treacy,fork}. In our experiment, we have used the grating pair
method to introduce negative dispersion to the idler photon.

The idler photon is first coupled into a 2 m long single-mode fiber
and the fiber polarization controller (FPC) is used to set the
polarization to maximize the diffraction efficiency. The grating
pair $G_1$ and $G_2$ and a plane mirror is used to setup the
negative dispersion device shown in Fig.~\ref{setup} \cite{grating}.
The distance between the two gratings is set at 10 cm and the
deviation angle at 750 nm is $8^\circ$. Since the grating $G_2$ is
not big enough to cover the full bandwidth, the grating pair system
slightly reduces the spectral bandwidth of the idler photon. After
experiencing negative dispersion, the idler photon is coupled into a
different 2 m long single-mode fiber for collimation and is sent
through a 1/2 m monochromator (CVI DK480) which functions as a
tunable narrowband filter. The monochromator M is used to
spectrally-resolve the entangled biphoton wave packet
\cite{baek08b,baek08c}.

Finally, the entangled biphoton wave packet is measured with two
single-photon detectors and a time-correlated single-photon counting
(TCSPC; PicoHarp 300, 16 ps resolution) device. The TCSPC histogram
directly visualizes the second-order Glauber correlation function
$G(t_1-t_2)$, if the observed effects are sufficiently bigger than
the resolution of the measurement system. The timing resolution of
the measurement system was found to be 762 ps which corresponds to
the width of the TCSPC histogram with the signal and the idler
photons coupled into 4 m long single-mode optical fibers.

\section{Results and Analysis}

\begin{figure}[t]
\centering
\includegraphics[width=3.2in]{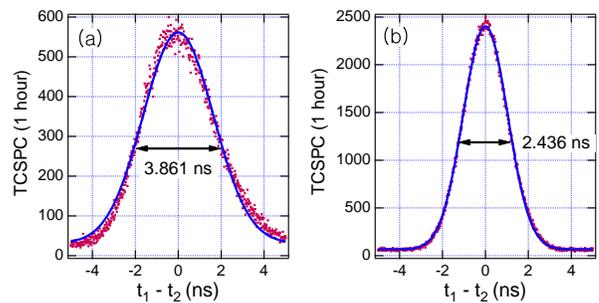}
\caption{Experimentally measured entangled photon wave packet.
Monochromator was not used for this measurement. (a) With positive
dispersion $\beta_1$ only, the wave packet has the FWHM width of
3.861 ns. (b) With both positive dispersion $\beta_1$ and negative
dispersion $\beta_2$, the wave acket has the FWHM width of 2.436 ns.
Solid lines are Gaussian fit to the data. }\label{data1}
\end{figure}

We first observed the entangled photon wave packet when the signal
photon was passed through a 1.6 km single-mode optical fiber which
introduces the positive dispersion $\beta_1$. The total length of
the idler photon's passage through the single-mode optical fiber is
4 m. Since we hope to observe the full TCSPC histogram which
represents the dispersion broadened entangled photon wave packet,
the monochromator M was not used for this measurement. The
experimental data are shown in Fig.~\ref{data1}(a). The measured
TCSPC histogram has the FWHM width of 3.861 ns which is
significantly bigger than the timing resolution of the measurement
system and is due to the dispersive broadening of the entangled
photon wave packet.

To experimentally demonstrate the nonlocal dispersion cancellation
effect, it is necessary to introduce negative dispersion $\beta_2$
to the idler photon so that the positive dispersion $\beta_1$
experienced by the signal photon is cancelled by the negative
dispersion $\beta_2$ experienced by the idler photon
\cite{franson,fitch}. The experiment was performed by directing the
idler photon through the grating pair system for negative
dispersion, while the signal photon's passage through the 1.6 km
single-mode fiber was not changed. As before, the monochromator M
was not used for this measurement as we aim to observe the full
TCSPC histogram.

The experimental data for this measurement are shown in
Fig.~\ref{data1}(b). While the data clearly demonstrate reduced FWHM
(2.436 ns) of the TCSPC histogram when compared to
Fig.~\ref{data1}(a), Fig.~\ref{data1} itself is quite insufficient
for a conclusive demonstration of the nonlocal dispersion effect. It
is because the grating pair system slightly reduces the spectral
bandwidth of the idler photon and the reduced FWHM shown in
Fig.~\ref{data1}(b) could be solely due to the bandwidth reduction.

It is thus  critically important to attribute how much of the wave
packet reduction observed in Fig.~\ref{data1} has actually come from
the nonlocal dispersion cancellation effect, if any. This critical
measurement was accomplished by spectrally resolving the two
entangled photon wave packets in Fig.~\ref{data1} by introducing the
monochromator M in the path of the idler photon
\cite{baek08b,baek08c}. The monochromator was set to have roughly
1.2 nm of passband and it functions as a wavelength-variable
bandpass filter. The TCSPC histograms measured for several values of
the wavelength setting of the monochromator correspond to spectrally
resolved components of the entangled photon wave packets shown in
Fig.~\ref{data1}.

Observation of nonlocal dispersion cancellation then requires
comparing the temporal spacing between the spectrally resolved
components of the entangled photon wave packets. If the temporal
spacing between the spectrally resolved components are reduced by
the introduction of the grating pair, it conclusively confirm the
nonlocal dispersion cancellation effect in
Ref.~\cite{franson,fitch}. On the other hand, if the temporal
separation remains the same while some of the components showing
reduced amplitudes, it would mean that the overall wave packet
reduction is actually due to the bandwidth filtering.

\begin{figure}[t]
\centering
\includegraphics[width=3.0in]{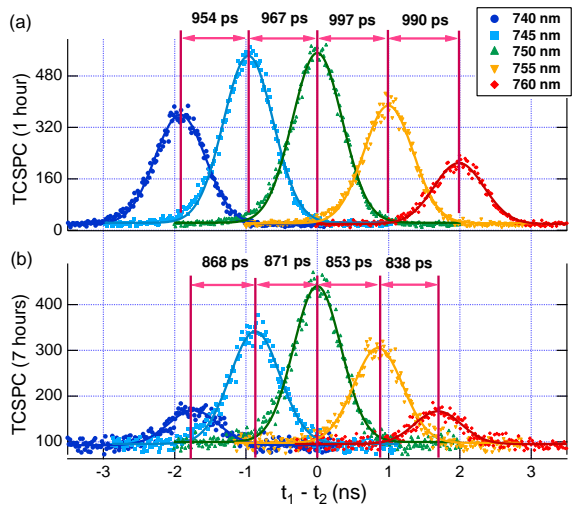}
\caption{Spectrally-resolved entangled photon wave packet with the monochromator. (a) With positive dispersion $\beta_1$ only. (b) With both positive dispersion $\beta_1$ and negative dispersion $\beta_2$. Solid lines are Gaussian fit to the data. The nonlocal dispersion cancellation effect is clear demonstrated. }\label{data2}
\end{figure}

The experimental data are shown in Fig.~\ref{data2}. In
Fig.~\ref{data2}(a), the entangled photon wave packet with positive
dispersion $\beta_1$ introduced in the path of the signal photon is
spectrally resolved. In Fig.~\ref{data2}(b), we show the spectrally
resolved entangled photon wave packet when both $\beta_1$ and
$\beta_2$ are introduced. Comparing Fig.~\ref{data2}(a) and
Fig.~\ref{data2}(b), we observe that the temporal spacings between
the spectrally resolved components are reduced when the idler photon
is subject to negative dispersion $\beta_2$. This is a clear and
conclusive signature of the nonlocal dispersion cancellation effect.
It is also interesting to observe that certain spectral components
(most notably, 740 nm and 760 nm) are strongly attenuated, causing
reduction of the entangled photon wave packet.

The above observations allow us to conclude that the reduced
entangled photon wave packet in Fig.~\ref{data1}(b) is due to both
the nonlocal dispersion cancellation effect and the bandwidth
reduction by the grating pair system. The experimental data in Fig.~\ref{data2} show that the reduction of the temporal spacing between 740 nm and 760 nm is 478 ps. To see what the theoretically expected value is, we need to make use of eq.~(\ref{eq8}). The introduction of $\beta_2$ reduces the width of $G^{(2)}$ function by
$$
\Delta t_{\textrm{NDC}} \approx 2\sqrt{\frac{2\ln2}{\gamma D^{2} L^{2}}} \beta_{2}z_{2}.
$$
In our experiment, $D L=88.9$ fs and $\beta_2 z_2$ for the grating pair system is given as \cite{diels},
$$\beta_{2}z_{2} = -\frac{\lambda_{2}} {2\pi c^2}\left(\frac{\lambda_{2}}{d}\right)\frac{G}{\cos(\theta^{\prime})^3},
$$
where $c$ is the vacuum speed of light, $\lambda_2= 750$ nm, $d=1/2400$ mm, $G = 10$ cm, and $\theta' = 60.45^\circ$. From these
values, we obtain $\beta_{2}z_{2} = -(2.03 \textrm{ ps})^2$. The theoretically calculated value of the reduction of the biphoton wave packet, therefore, is approximately 496 ps. Considering the measurement errors for evaluating $\beta_2 z_2$, our experimental observation agrees very well with the theoretical prediction.

We therefore estimate that, out of 1.425 ns reduction of the wave
packet shown in Fig.~\ref{data1}, roughly 478 ps comes from the
nonlocal dispersion cancellation effect and the rest comes from the
bandwidth filtering at the grating pair. To improve the nonlocal
dispersion cancellation effect, it is necessary to eliminate the
bandwidth reduction effect of the grating pair system and it can be
done by replacing the grating $G_2$ with a larger one.

\section{Discussion}

Using the spontaneous parametric down-conversion photon pairs whose central frequencies are anticorrelated
as in eq.~(\ref{eq33}), we have shown that two-photon entanglement (i.e., the coherent superposition of frequency-anticorrelated biphoton amplitudes) can exhibit the nonlocal
dispersion cancellation effect, see eq.~(\ref{eq8}) \cite{franson}. Naturally, one might wonder whether some classical
frequency-anticorrelated states could also exhibit the nonlocal dispersion cancellation effect. In this section, we discuss these classical cases and show that it is not possible to achieve the nonlocal dispersion cancellation effect of Ref.~\cite{franson} using classical states.

\subsection{A frequency-anticorrelated mixture of two-photon states}

Spontaneous parametric down-conversion produces a two-photon pure
state as shown in eq.~(\ref{eq33}). Let us see what would happen if
phase coherence is removed so that the input state is now a
frequency-anticorrelated mixture of two-photon states given as
\begin{eqnarray}\label{eq88}
\nonumber\rho_{mixed}&=&\int d\nu
f_{1}(\nu)f_{2}(\nu)a^{\dagger}(\Omega_{1}+\nu)a^{\dagger}(\Omega_{2}-\nu)|0\rangle
\langle 0| \\
&&a(\Omega_{1}+\nu)a(\Omega_{2}-\nu),
\end{eqnarray}
where $f_{1}(\nu)=f_{2}(\nu)=e^{-\nu^2/2\sigma^2}$ are spectral
amplitudes of photon 1 and photon 2 and we assume they follow
gaussian spectral distribution. The normalization condition gives $\int d\nu f_{1}(\nu)f_{2}(\nu)=1$. Such states might experimentally be approximated by using a pair of strongly attenuated cw lasers whose frequencies are tuned in opposite directions by the random amount $\nu$ and filtered by Gaussian filters with transmission function  $f_{1}(\nu)$ and $f_{2}(\nu)$.

When $\rho_{mixed}$ is used in the experimental setup shown in Fig.~\ref{idea}, the joint detection rate between two detectors $D_1$ and $D_2$ is proportional to the Glauber 2nd-order correlation function
\begin{equation}\label{eq888}
G^{(2)}(t_1,t_2) =
Tr[\rho_{mixed}E_{1}^{(-)}(t_{1})E_{2}^{(-)}(t_{2})E_{2}^{(+)}(t_{2})E_{1}^{(+)}(t_{1})],
\end{equation}
where $E_{1}^{(+)}(t_{1})= \int d \nu
a(\nu)e^{i\{k_{1}z_{1}-(\Omega_{1}+\nu) t_{1}\}}$, the positive
frequency component of the electric field operator at detector $D_1$, and
$E_2^{(+)}$ is defined similarly. By combining eq.~(\ref{eq88}) and eq.~(\ref{eq888}), we finally
obtain
\begin{equation}
G^{(2)}(t_1,t_2)=1.
\end{equation}
Physically, this means that the joint detection probabilities of two
detectors do not depend on time at all. In other words, even $D_{1}$
detects a photon at $t_1$, we have no information about when $D_2$
would click. The joint detection probability does not show timing
information at all and, therefore, is meaningless to discuss the
dispersive effect with the two-photon mixed state. This result tells
us that coherence between two-photon probability amplitudes (i.e.,
entanglement) is essential for the nonlocal dispersion cancellation
effect.

\subsection{A classical pulse pair}

Let us now consider whether the nonlocal dispersion cancellation
effect of Ref.~\cite{franson} can be exhibited by a classical pulse
pair. First, assume two ultrafast pulses which are initially
coincident in time \cite{franson}. The frequency of each pulse can
be defined as $\omega_1=\Omega_1+\nu_1$ and
$\omega_2=\Omega_2-\nu_2$, respectively. Here $\nu_i$ is the
detuning frequency from the central frequency $\Omega_i$ and the
subscript $i$ labels each pulse ($i=1,2$). The wave number of the
photon can then be expressed as $k_{i}(\Omega_{i} \pm \nu_i)
\thickapprox k_{i}(\Omega_{i}) \pm \alpha_{i}\nu_i
+\beta_{i}\nu_i^{2}$. Note that  $\alpha$ and $\beta$ are the
first-order and the second-order dispersion which are responsible
for the wave packet delay and the wave packet broadening,
respectively.

After pulse 1 has propagated through the dispersive medium
$\beta_1$, see Fig.~\ref{idea}, the electric field at the detector
can then be written as
\begin{eqnarray}\label{eq9}
\nonumber E_{1}(t_1,z_1)&=& \int_{-\infty}^{\infty}d \nu_1
\frac{E_0}{2 \pi}e^{-\nu_1^2/2\sigma_0^2}
e^{i(k_{1}(\Omega_1)+\alpha_1
\nu_1+\beta_1 \nu_1^2) z_1} \\
&&\times e^{-i(\Omega_1+\nu_1) t_{1}},
\end{eqnarray}
where $E_0$ is a constant,  $\sigma_0$ is the bandwidth of the
pulse, $z_1$ is the distance between the light source and the
detector, and $t_1$ is the detection time. $E_{2}(t_2,z_2)$ for
pulse 2 is defined similarly.

The intensity detected at $D_1$ is $I_1(z_1,t_1)=|E_1(z_1,t_1)|^2$ and is calculated to be
\begin{equation}\label{eq10}
I_1(z_1, t_1)=\frac{E_0^2}{4 \pi
|a_1|^{2}}e^{-(\alpha_{1}z_{1}-t_{1})^2/2 \sigma_{1}^2},
\end{equation}
where $a_1^2=1/2\sigma_0^2-i\beta_{1}z_{1}$ and
$\sigma_{1}^2=2\sigma_{0}^2(1/4\sigma_{0}^4+\beta_{1}^{2}z_{1}^{2})$. The intensity at $D_2$ is calculated similarly as
\begin{equation}\label{eq11}
I_2(z_2, t_2)=\frac{E_0^2}{4 \pi
|a_2|^{2}}e^{-(\alpha_{2}z_{2}-t_{2})^2/2 \sigma_{2}^2},
\end{equation}
where $a_2^2=1/2\sigma_0^2-i\beta_{2}z_{2}$ and
$\sigma_{2}^2=2\sigma_{0}^2(1/4\sigma_{0}^4+\beta_{2}^{2}z_{2}^{2})$.

The joint detection probability, i.e., the probability that two detectors $D_1$ and $D_2$ click simultaneously at
times $t_1$ and $t_2=t_{1}+\tau$ is then given as $\eta
I_{1}(z_1,t_1)I_{2}(z_2,t_1+\tau)$, where $\eta$ is a constant
related to the detection efficiency, and is calculated to be
\begin{equation}\label{eq12}
P(\tau)= \int_{-\infty}^{\infty}d t_1 \eta
I_{1}(z_1,t_1)I_{2}(z_2,t_1+\tau)=C
e^{-\frac{(\tau-\overline{\tau})^2}{2(\sigma_{1}^{2}+\sigma_{2}^{2})}},
\end{equation}
where $\overline{\tau}=\alpha_{2}z_{2}-\alpha_{1}z_{1}$ and $C=\eta
E_{0}^4/[(4
\pi)^{2}|a_{1}|^{2}|a_{2}|^{2}]\int_{-\infty}^{\infty}\exp[-\frac{1}{2}(\frac{1}{\sigma_{1}^2}+\frac{1}{\sigma_{2}^2})t_{1}^{2}]dt_{1}$.

The joint detection or the coincidence distribution $P(\tau)$ then has a total width
\begin{equation}\label{eq13}
\sigma_{T}^{2}=\sigma_{1}^{2}+\sigma_{2}^{2}=2\sigma_{0}^{2}(1/2\sigma_{0}^{4}+\beta_{1}^{2}z_{1}^{2}+\beta_{2}^{2}z_{2}^{2}),
\end{equation}
and for large distances and dispersions, eq.~(\ref{eq13})  becomes
\begin{equation}\label{eq14}
\sigma_{T}^{2}\approx
2\sigma_{0}^{2}(\beta_{1}^{2}z_{1}^{2}+\beta_{2}^{2}z_{2}^{2}).
\end{equation}

Finally, the full width at half maximum value of the temporal width
of $P(\tau)$ is given as
\begin{equation}\label{eq15}
\Delta t
\approx4\sqrt{\ln2}\sigma_{0}\sqrt{\beta_{1}^{2}z_{1}^{2}+\beta_{2}^{2}z_{2}^{2}}.
\end{equation}
Since eq.~(\ref{eq15}) depends on the summation of squares of
individual dispersion values, nonlocal dispersion cancellation  is
not possible using two classical pulses even in the case
$\beta_1=-\beta_2$. Clearly, this is due to the local nature of
classical fields \cite{franson}. Even if the two classical pulses
are anticorrelated in their central frequencies, the rest of
frequency components of pulse 1 and pulse 2 do not have spectral nor
temporal correlations. As a result, the local dispersion experienced
by pulse 1 cannot be canceled by the local dispersion experienced by
pulse 2.

\begin{figure}[t]
\centering
\includegraphics[width=2.9in]{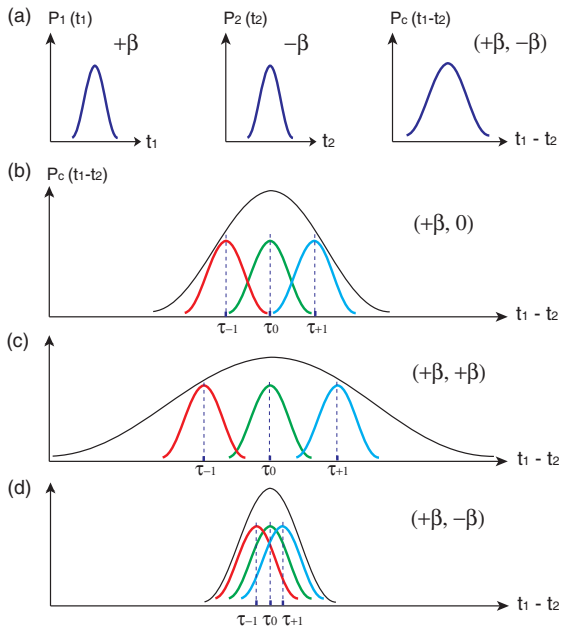}
\caption{The joint detection probability for a number of cases
involving classical pulse pairs under group velocity dispersion. (a)
Detection probabilities for pulse 1, $P_1(t_1)$, and pulse 2,
$P_2(t_2)$, after they have propagated through $+\beta$ and $-\beta$
media, respectively. The joint detection probability $P_c(t_1-t_2)$
is always broader then the single detection probabilities. Note that
($+\beta$,$-\beta$) refers to the case in which pulse 1 (pulse 2)
goes through the positive (the negative) dispersive medium. Figures
(b), (c), and (d) show the joint detection probabilities for a
frequency-anticorrelated mixture of classical pulse pairs. Note that
($+\beta$,$0$) refers to the case in which pulse 1 goes through the
positive dispersion medium while pulse 2 goes through a
non-dispersive medium. ($+\beta$,$+\beta$), and ($+\beta$,$-\beta$)
have similar meanings. If spectrally identical sources are used, the
the overall joint detection probabilities cannot be smaller than the
individual probabilities. See text for details. }\label{pulse}
\end{figure}

The effect of dispersion on the joint detection probability of the
classical pulse pair can be easily described using
Fig.~\ref{pulse}(a).  The detection probability as a function of
time for a short pulse (pulse 1) after it has propagated through a
positive dispersion medium $+\beta$ is given as $P_{1}(t_{1})$. The
detection probability for pulse 2 after it has propagated through a
negative dispersion medium $-\beta$ is given as $P_{2}(t_{2})$.
$P_{1}(t_{1})$ and $P_{2}(t_{2})$ show the wave packet broadening
due to eq.~(\ref{eq10}) and eq.~(\ref{eq11}), respectively. The
result in eq.~(\ref{eq15}) shows that the joint detection
probability $P_{c}(t_{1}-t_{2})$ would then be much broader than
$P_{1}(t_{1})$ and $P_{2}(t_{2})$ individually as depicted in
Fig.~\ref{pulse}(a). Note that this case is referred in
Fig.~\ref{pulse}(a) as  ($+\beta$,$-\beta$). Exactly the same
results are obtained for other cases, namely,  ($+\beta$, $+\beta$),
($-\beta$, $+\beta$), and ($-\beta$,$-\beta$).

\subsection{A frequency-anticorrelated mixture of classical pulse pairs}

Let us now consider a frequency-anticorrelated mixture of classical
pulse pairs. Here we imagine many frequency-anticorrelated pulse
pairs, each with a random amount of detuning $\nu$ from the central
frequency $\Omega$. Thus, a pulse pair is always
frequency-anticorrelated but no pairs have the same amount of
detuning. We also assume a sufficiently broadband Gaussian filters
are placed in front of the detectors.

Clearly, the joint detection probability for each pulse pair is
described by eq.~(\ref{eq15}). However, since each pulse pair has
different central frequencies than the previous/next ones, the
coincidence peak for each pulse pair, after they have propagated
through the dispersion media, would then appear at different times
due to the dispersion.

The coincidence peak corresponding to the $m$'th pulse pair occurs at
\begin{equation}\label{eq16}
\tau_{m}=\frac{z_1}{v_{1}(\Omega+m \nu)}-\frac{z_2}{v_{2}(\Omega-m
\nu)},
\end{equation}
where $m$ is an integer, $v_i$ is the group velocity of pulse $i$
($i=1,2$), and $z_i$ is the distance between the source and the
detector. Note that $\tau_m$, the time at which the coincidence peak corresponding to the $m$'th pulse pair occurs, is simply the relative group delay between the two pulses. In what follows, we will explore how the statistical distribution of $\tau_m$ is affected by the dispersive media.

\textbf{CASE I} \textbf{($+\beta$, 0)}: Consider the case that pulse
1 is sent to a positive dispersive medium, $+\beta$, and pulse 2 is
sent to a nondispersive medium (e.g., air). In this case, the group
velocity of pulse 2 can be approximated as
$v_{2}(\Omega-m\nu)\approx c$, where $c$ is the speed of light in
the air, and eq.~(\ref{eq16}) can be rewritten as $\tau_{m} =
z_1/v_{1}(\Omega+m \nu)-z_2/c$. When the cental frequency of pulse 1
is tuned to $\Omega \pm \nu$ ($\nu > 0$) from $\Omega$, the
coincidence peak is shifted from $\tau_0 = z_1/v_{1}(\Omega)-z_2/c$
to $\tau_{\pm1} = z_1 / v_{1}(\Omega \pm \nu)-z_2/c$. Since the
group delay of pulse 1 monotonously increases as the frequency
increases for the positive dispersion medium, the group delays
corresponding to the three coincidence peaks satisfy the relation,
$\tau_{-1}< \tau_{0} < \tau_{+1}$, as shown in Fig.~\ref{pulse} (b).

We can then define the measure of the width of the coincidence
distribution as

\begin{equation}\label{eq17}
\Delta \tau (+\beta, 0)   \equiv
\tau_{+1}-\tau_{0}=\frac{z_1}{v_{1}(\Omega+\nu)}-\frac{z_1}{v_{1}(\Omega)}.
\end{equation}
When the coincidence measurement is done for a sufficiently long
time, it will take the form of a broadened Gaussian envelope (due to
the Gaussian filters) as shown in Fig.~\ref{pulse}(b).

\textbf{CASE II} \textbf{($+\beta$, $+\beta$)}: Let us now consider
the case that pulse 1 and pulse 2 are both sent to positive
dispersion media. Considering that the pulse pair is
frequency-anticorrelated in their central frequencies and the
dispersion is positive for both pulses, the relative group delays
can be calculated using eq.~(\ref{eq16}) as
$\tau_{0}=z_1/v_{1}(\Omega)-z_2/v_{2}(\Omega)$ and
$\tau_{+1}=z_1/v_{1}(\Omega+\nu)-z_2/v_{2}(\Omega- \nu)$. The
coincidence width is then calculated to be

\begin{equation}\label{eq18}
\Delta \tau (+\beta, +\beta)=\tau_{+1}-\tau_{0}=\Delta \tau (+\beta,
0)+ \frac{z_2}{v_{2}(\Omega)}-\frac{z_2}{v_{2}(\Omega-\nu)},
\end{equation}
where $\Delta \tau (+\beta, 0)$ is defined in eq.~(\ref{eq17}).
Since pulse 2 propagates through the positive dispersion medium,
$v_{2}(\Omega) < v_{2}(\Omega-\nu)$ and, therefore, $\Delta \tau
(+\beta, +\beta)> \Delta \tau (+\beta, 0)$. This case, shown in
Fig.~\ref{pulse}(c), therefore gives broader coincidence
distribution than the $(+\beta, 0)$ case shown in
Fig.~\ref{pulse}(b).


\textbf{CASE III} \textbf{($+\beta$, $-\beta$)}: Finally we discuss
the case in which pulse 1 propagates through the positive dispersion
medium while pulse 2 propagates through the negative dispersion
medium. Remember that the pulse pair mixture is
frequency-anticorrelated and each pair has random detuning. Since
pulse 2 propagates through the negative dispersion medium,
$v_{2}(\Omega) > v_{2}(\Omega-\nu)$ and, therefore, $\Delta \tau
(+\beta, -\beta)=\tau_{+1}-\tau_{0}=\Delta \tau (+\beta,
0)+z_2/v_{2}(\Omega)-z_2/v_{2}(\Omega-\nu) < \Delta \tau (+\beta,
0)$. As a result, the relative group delay $\tau_m$ does not
increase much from the $\tau_{0}$ value even with large $m$'s. The
coincidence peak, therefore,  occurs mostly at around
$\tau_0=z_1/v_g(\Omega)-z_2/v_g(\Omega)$ as shown in
Fig.~\ref{pulse}(d).

Thus, for the frequency-anticorrelated mixture of classical pulse
pairs, we arrive at the relation,
\begin{equation}\label{eq19}
\Delta \tau (+\beta, +\beta)> \Delta \tau (+\beta, 0)> \Delta \tau
(+\beta, -\beta).
\end{equation}
One might try to interpret the above result as showing, even with
the frequency-anticorrelated mixture of classical pulse pairs, it is
possible to achieve the nonlocal dispersion cancellation effect.
This, however, is an erroneous conclusion because the minimum
coincidence distribution in time is limited by eq.~(\ref{eq15}),
depicted in Fig.~\ref{pulse}(a), which is quite different from the
quantum nonlocal dispersion cancellation effect in eq.~(\ref{eq8}).

\section{Conclusion}

We have reported an experimental demonstration of the quantum
nonlocal dispersion cancellation effect of Ref.~\cite{franson} in
which the signal photon is subject to positive dispersion while its
entangled twin photon, remotely located, is subject to negative
dispersion. In this work, for the first time, we have explicitly
demonstrated the narrowing of the joint detection probability
function by spectrally resolving the dispersion-broadened two-photon
wave packet and the amount of narrowing has shown to be consistent
with the theoretical calculations based on the quantum nonlocal
dispersion cancellation effect. We have also shown theoretically
that classical states, e.g., a two-photon mixed state and a
frequency-anticorrelated mixture of classical pulse pairs, cannot
demonstrate the nonlocal dispersion cancellation effect. We expect
that the nonlocal dispersion cancellation effect will play an
important role in fiber-based quantum communication and quantum
metrology applications as the method to remotely remove/compensate
unwanted dispersive effects on nonclassical wave packets.

\section*{Acknowledgments}

This work was supported, in part, by the Korea Research Foundation
(KRF-2006-312-C00551) and the Korea Science and Engineering
Foundation (R01-2006-000-10354-0), and the Ministry of Knowledge and
Economy of Korea through the Ultrafast Quantum Beam Facility
Program.



\begin{thebibliography}{99}
\bibitem{gis2} The demonstration in Ref.~\cite{gis} is local in the sense that both photons propagate through the same optical fiber. Furthermore, actual reduction of a broadened wave packet is not experimentally demonstrated.
\bibitem{note0} We have assumed $\gamma^2 D^4 L^4  \ll (\beta_1 z_1 + \beta_2 z_2)^2$. This condition holds in general as the crystal parameter $DL$ is much smaller than the external parameters $\beta_1 z_1$ and $\beta_2 z_2$.
\end{thebibliography}
\end{document}